\documentclass[amsmath,aps,showpacs,a4paper,10pt]{revtex4}

 \usepackage{epsf}
 \usepackage{graphicx}    

 \usepackage{amssymb}

 \usepackage{enumerate}   

\begin{document}

 \newcommand{\be}[1]{\begin{equation}\label{#1}}
 \newcommand{\ee}{\end{equation}}
 \newcommand{\bea}{\begin{eqnarray}}
 \newcommand{\eea}{\end{eqnarray}}
 \def\disp{\displaystyle}

 \def\gsim{ \lower .75ex \hbox{$\sim$} \llap{\raise .27ex \hbox{$>$}} }
 \def\lsim{ \lower .75ex \hbox{$\sim$} \llap{\raise .27ex \hbox{$<$}} }

 \begin{titlepage}

 \begin{flushright}
 arXiv:1907.12517
 \end{flushright}

 \title{\Large \bf Lema\^{\i}tre-Tolman-Bondi
 Static Universe\\ in Rastall-like Gravity}

 \author{Zhong-Xi~Yu\,$^{1,}$\footnote{email address:\ 547410406@qq.com}\,,
 Shou-Long~Li\,$^{2,}$\footnote{email address:\ shoulongli@hunnu.edu.cn}\,,
 Hao~Wei\,$^{1,}$\footnote{Corresponding author;\ email
 address:\ haowei@bit.edu.cn}}
 \affiliation{\vspace{1.8mm}
 $^{1}$School of Physics, Beijing Institute of
 Technology, Beijing 100081, China \vspace{1mm}\\
 $^{2}$Department of Physics and Synergetic Innovation
 Center for Quantum Effect and Applications, Hunan Normal
 University, Changsha 410081, China}

 \begin{abstract}\vspace{1cm}
 \centerline{\bf ABSTRACT}\vspace{2mm}
 In this work, we try to obtain a stable Lema\^{\i}tre-Tolman-Bondi (LTB)
 static universe, which is spherically symmetric and radially inhomogeneous.
 However, this is not an easy task, and fails in general relativity (GR) and
 various modified gravity theories, because the corresponding LTB static
 universes must reduce to the Friedmann-Robertson-Walker (FRW) static
 universes. We find a way out in a new type of modified gravity theory,
 in which the conservation of energy and momentum is broken. In this work,
 we have proposed a novel modification to the original Rastall gravity. In
 some sense, our Rastall-like gravity is essentially different from GR and
 the original Rastall gravity. In this Rastall-like gravity, LTB static
 solutions have been found. The stability of LTB static universe against
 both the homogeneous and the inhomogeneous scalar perturbations is also
 discussed in details. We show that a LTB static universe can be stable
 in this Rastall-like gravity.
 \end{abstract}


 \maketitle

 \end{titlepage}

 \renewcommand{\baselinestretch}{1.0}


\section{Introduction}\label{sec1}

As is well known, modern cosmology began with the application
 of general relativity (GR) to the universe, soon after the
 birth of GR. The first cosmological model developed by
 Einstein himself in 1917~\cite{Einstein:1917ce} is the
 well-known Einstein static universe, which is homogeneous,
 isotropic, and spatially closed. It can be static by the
 help of a positive cosmological constant counteracting the
 attractive effects of gravity on ordinary matter. However,
 in 1929, Hubble found that the universe is expanding (rather
 than static), by examining the relation between distance and
 redshift of galaxies~\cite{Hubble:1929ig}. On the other hand,
 in 1930, Eddington~\cite{Eddington:1930zz} argued that
 the Einstein static universe is unstable with respect to
 homogeneous and isotropic scalar perturbations in GR, and
 hence the universe cannot be static in the presence of
 perturbations. Therefore, Einstein abandoned the idea of
 static universe (and the cosmological constant
 as the ``\,biggest blunder\,'' in his life~\cite{Gamov:1970}).

Recently, the Einstein static universe has been revived to
 avoid the big bang singularity in the emergent universe
 scenario~\cite{Ellis:2002we,Ellis:2003qz}. In such kind of
 scenario, the Einstein static universe is the initial state
 for a past-eternal inflationary cosmological model and then
 evolves to an inflationary era. So, there is no big bang
 singularity, and no exotic physics is involved. The quantum
 gravity regime can even be avoided, if the size of Einstein
 static universe is big enough. In fact, it is argued that the
 Einstein static state is favored by entropy considerations as
 the initial state for the universe~\cite{Gibbons:1987jt,
 Gibbons:1988bm,Mulryne:2005ef}.

Motivated by the emergent universe scenario, in the past
 decade, the Einstein static universe was extensively studied
 in many gravity theories. It has been reconsidered in GR,
 and the Einstein static universe can be stable against
 perturbations, if the universe contains a perfect fluid with
 $c_s^2>1/5$~\cite{Gibbons:1987jt,Gibbons:1988bm,Barrow:2003ni}. It
 is also very interesting to consider the Einstein static
 universe in various modified gravity theories, for example,
 loop quantum cosmology~\cite{Parisi:2007kv},
 $f(R)$ theory~\cite{Boehmer:2007tr,Seahra:2009ft,Goheer:2008tn}, $f(T)$
 theory~\cite{Wu:2011xa,Li:2013xea}, modified Gauss-Bonnet
 gravity ($f(G)$ theory)~\cite{Bohmer:2009fc,Huang:2015kca}, Brans-Dicke
 theory~\cite{delCampo:2007mp,delCampo:2009kp,Huang:2014fia,Miao:2016obc},
 Horava-Lifshitz theory~\cite{Wu:2009ah,Boehmer:2009yz,Heydarzade:2015hra},
 massive gravity~\cite{Parisi:2012cg,Mousavi:2016eof},
 braneworld scenario~\cite{Gergely:2001tn,Zhang:2010qwa,Atazadeh:2014xsa},
 Einstein-Cartan theory~\cite{Atazadeh:2014ysa}, $f(R, T)$
 gravity~\cite{Shabani:2016dhj}, hybrid metric-Palatini
 gravity~\cite{Boehmer:2013oxa}, Eddington-inspired Born-Infeld
 theory~\cite{Li:2017ttl}, degenerate massive gravity~\cite{Li:2019laq},
 and so on~\cite{Boehmer:2003iv,Goswami:2008fs,Canonico:2010fd,
 Zhang:2016obw,Carneiro:2009et,Boehmer:2015ina,
 Atazadeh:2015zma,Atazadeh:2016yeh}.

We note that (almost) all relevant works in the literature by
 now only considered the Einstein static universe, which is
 homogeneous and isotropic, and hence is described by a
 Friedmann-Robertson-Walker (FRW) metric. In other words, they
 assumed the cosmological principle.

However, as a tenet, the cosmological principle is not born to
 be true. Actually, this assumption has not yet been well proven on
 cosmic scales $\gtrsim 1\,{\rm Gpc}$~\cite{Caldwell:2007yu}.
 Therefore, it is still of interest to test both the homogeneity and
 the isotropy of the universe carefully. In fact, they could
 be violated in some theoretical models, such as
 Lema\^{\i}tre-Tolman-Bondi (LTB) model~\cite{LTB} (see also
 e.g.~\cite{Yan:2014eca,Deng:2018yhb,Deng:2018jrp,Qiang:2019zrs} and
 references therein) violating the cosmic homogeneity, and the
 exotic G\"odel universe~\cite{Godel:1949ga} (see also
 e.g.~\cite{Li:2016nnn} and references therein), most of the
 Bianchi type I$\,\sim\,$IX universes~\cite{Bianchi}, Finsler
 universe~\cite{Li:2015uda}, violating the cosmic isotropy. On
 the other hand, many observational hints of the cosmic
 inhomogeneity and/or anisotropy have been claimed in the
 literature (see e.g.~\cite{Deng:2018yhb,Deng:2018jrp,Qiang:2019zrs}
 for brief reviews), including type Ia supernovae (SNIa),
 cosmic microwave background (CMB), baryon acoustic oscillations (BAO),
 gamma-ray bursts (GRBs), integrated Sachs-Wolfe effect,
 Sunyaev-Zel'dovich effect, quasars, radio galaxies, and so on.
 Therefore, on both the theoretical and the observational sides, it
 is reasonable to consider the cosmological models violating
 the cosmological principle.

It is natural to ask ``\,why must the initial state for the universe
 be homogeneous and isotropic?\,'' If the initial state is random, it
 has great probability to be inhomogeneous and/or anisotropic.
 Therefore, it is interesting to consider a static universe violating
 the cosmological principle.

In the present work, we are interested in the well-known
 LTB model~\cite{LTB} (see also e.g.~\cite{Yan:2014eca,Deng:2018yhb,
 Deng:2018jrp,Qiang:2019zrs} and references therein). In this
 model, the universe is spherically symmetric and radially
 inhomogeneous, and we are living in a locally underdense void
 centered nearby our location. As is well known, without
 invoking dark energy or modified gravity, it is possible to explain
 the apparent cosmic acceleration discovered in 1998 by using
 the LTB model~\cite{Alnes:2005rw,GarciaBellido:2008nz,
 Enqvist:2006cg,Celerier:2012xr,Ishak:2013vha,Clifton:2008hv,
 Zhang:2012qr,Yan:2014eca,Yousaf:2017xec}. This fact further
 justifies the motivation to consider a LTB static universe.

However, it is not an easy task to obtain a stable LTB static universe in
 GR and various modified gravity theories. We will briefly discuss this
 issue in Sec.~\ref{sec2}. Knowing the cause of failure, we find a way out
 in a new type of modified gravity theory, namely Rastall-like gravity,
 which will be briefly introduced in Sec.~\ref{sec3}. Then, in
 Secs.~\ref{sec4} and \ref{sec5}, we obtain the LTB static solutions in
 Rastall-like gravity without and with a cosmological constant,
 respectively. The stability of LTB static universe against both the
 homogeneous and the inhomogeneous scalar perturbations is also discussed
 in details. In Sec.~\ref{sec6}, some brief concluding remarks are given.


\section{The failure of LTB static universe
 in GR and various~modified~gravity~theories}\label{sec2}

The LTB metric, in comoving coordinates ($r$, $\theta$, $\phi$) and
 synchronous time $t$, is given by~\cite{LTB,Alnes:2005rw,
 GarciaBellido:2008nz,Enqvist:2006cg,Celerier:2012xr,Ishak:2013vha,
 Clifton:2008hv,Zhang:2012qr,Yan:2014eca}
 \be{eq1}
 ds^2 = -dt^2 + \frac{A^{\prime\,2}(r,t)}{1-K(r)}\,dr^2
 +A^2(r,t)\,d\Omega^2\,,
 \ee
 where $d\Omega^2=d\theta^2+\sin^2\theta\,d\phi^2$, and a prime
 denotes a derivative with respect to $r$. $K(r)$ is an
 arbitrary function of $r$, playing the role of spatial
 curvature. In general, $A(r,t)$ is an arbitrary function of
 $r$ and $t$, playing the role related to the scale factor.
 Obviously, the LTB metric reduces to the well-known FRW
 metric if $A(r,t)=a(t)\,r$ and $K(r)=Kr^2$.


Let us consider the LTB static universe, in which $A=A_0(r)$ is
 independent of the time $t$, and the subscript ``\,0\,'' indicates
 the quantities related to the static solutions. In this case,
 $\dot{A}=0$ and $\ddot{A}=0$, where a dot denotes a derivative
 with respect to $t$. Now, by definition, the Einstein
 tensor is given by
 \be{eq2}
 G^\nu{}_\mu=R^\nu{}_\mu-\frac{1}{2}\,R\,\delta^\nu{}_\mu=
 {\rm diag}\left(-\frac{K_0}{A_0^2}-\frac{K_0^\prime}{A_0 A_0^\prime}\,,\,
 -\frac{K_0}{A_0^2}\,,\,
 -\frac{K_0^\prime}{2A_0 A_0^\prime}\,,\,
 -\frac{K_0^\prime}{2A_0 A_0^\prime}
 \right),
 \ee
 where $\delta^\nu{}_\mu={\rm diag}\left(1,\,1,\,1,\,1\right)$,
 $R_{\mu\nu}$ is the Ricci tensor, $R=R^\mu{}_\mu$, and the
 Greek indices $\mu$, $\nu$ run over 0, 1, 2, 3. We note that
 the last three diagonal components $G^1{}_1\not=G^2{}_2=G^3{}_3$ in
 general. On the other hand, if the universe contains a
 perfect fluid, the corresponding energy-momentum tensor reads
 \be{eq3}
 T^{\mu\nu}=(\rho+p)\,U^\mu U^\nu+p\,g^{\mu\nu}\,,
 \ee
 or equivalently
 \be{eq4}
 T^\nu{}_\mu={\rm diag}\left(-\rho,\,p,\,p,\,p\right)\,,
 \ee
 where 4-velocity $U^\mu=dx^\mu/d\tau$ satisfies
 $g_{\mu\nu}U^\mu U^\nu=-1$, while $\tau$ is the proper time,
 $\rho$ and $p$ are energy density and pressure, respectively. In a
 proper (comoving) frame, $U^\mu=(1,\,0,\,0,\,0)$. In general, $\rho$
 and $p$ are functions of $r$ and $t$. In the case of LTB static universe,
 $\rho=\rho_0(r)$ and $p=p_0(r)$ are independent of $t$.

In GR, the field equations are
 $G^\nu{}_\mu=8\pi G_N T^\nu{}_\mu$, where $G_N$ is the
 Newtonian gravitational constant. Since the last three
 diagonal components of $T^\nu{}_\mu$ are equal, we must have
 $G^1{}_1=G^2{}_2=G^3{}_3=8\pi G_N p\,$. In the case of
 LTB static universe, from Eq.~(\ref{eq2}), it requires
 \be{eq5}
 \frac{K_0}{A_0^2}=
 \frac{K_0^\prime}{2A_0 A_0^\prime}=\frac{dK_0}{dA_0^2}\,,
 \ee
 which means $K_0/A_0^2={\cal K}={\rm const.}$ Introducing
 $a_0=\left|{\cal K}\right|^{-1/2}$, $\tilde{r}=A_0(r)/a_0$,
 $\tilde{\cal K}={\cal K}a_0^2=\pm 1$ if ${\cal K}\not=0$,
 or introducing $\tilde{r}=A_0(r)/a_0$, $\tilde{\cal K}=0$,
 $a_0={\rm const.}\not=0$ if ${\cal K}=0$, the metric of
 LTB static universe becomes
 \be{eq6}
 ds^2 = -dt^2 +a_0^2\left(\frac{d\tilde{r}^2}{1-
 \tilde{\cal K}\tilde{r}^2}+\tilde{r}^2\,d\Omega^2\right)\,,
 \ee
 which is nothing but the one of FRW static universe. So, the
 LTB static universe in GR fails. In addition, we have
 $8\pi G_N p_0=G^1{}_1=-K_0/A_0^2=-{\cal K}={\rm const.}$,
 which is also independent of $r$. Actually, one
 can find $\partial_r p_0=0$ from the conservation equations
 $T^\nu{}_{\mu;\nu}=0$. In fact, this gives an instructive hint
 for the failure of LTB static universe in GR.

Furthermore, we have also considered the LTB static universes in
 various modified gravity theories, such as $f(R)$ theory,
 $f(T)$ theory, Brans-Dicke theory, modified Gauss-Bonnet
 gravity ($f(G)$ theory), and they all failed because the
 corresponding LTB static universes all reduced to the FRW
 static universes. In fact, one can always recast the field
 equations of modified gravity theory as the form of
 \be{eq7}
 G^\nu{}_\mu+M^\nu{}_\mu=8\pi G_N T^\nu{}_\mu\,,
 \ee
 where $M^\nu{}_\mu$ is the modification term with respect to GR. In
 the case of LTB static universe, if we require that the conservation
 equations $T^\nu{}_{\mu;\nu}=0$ hold (which are actually equivalent to
 $M^\nu{}_{\mu;\nu}=0$ because $G^\nu{}_{\mu;\nu}=0$ always), the last
 three diagonal components of $M^\nu{}_\mu$ should be equal in various
 modified gravity theories such as $f(R)$ theory, $f(T)$ theory, Brans-Dicke
 theory, and modified Gauss-Bonnet gravity ($f(G)$ theory). Since the last
 three diagonal components of $T^\nu{}_\mu$ are also equal, it is necessary
 to require $G^1{}_1=G^2{}_2=G^3{}_3$. Following the similar derivations
 in GR, the LTB static universes in these modified gravity theories also
 fail, because they must reduce to the FRW static universes.


\section{The Rastall-like gravity theory}\label{sec3}


\subsection{The original Rastall gravity theory}\label{sec3a}

From the discussions in Sec.~\ref{sec2}, it is easy to see that
 the conservation equations $T^\nu{}_{\mu;\nu}=0$ might be
 responsible for the failure of LTB static universe in GR and
 various modified gravity theories. In order to obtain a
 successful LTB static universe, a possible way out might be
 breaking the conservation equations $T^\nu{}_{\mu;\nu}=0$.
 In the literature, there exists some models breaking the
 conservation of energy and momentum. For instance, the famous
 steady state model of the expanding universe proposed by
 Hoyle, Bondi and Gold in 1948~\cite{Hoyle:1948zz,Bondi:1948qk}
 requires continuous creation of matter (particle creation)
 from nothing. In fact, it is not completely unacceptable to
 consider such kind of models.

In this work, we are interested in the so-called Rastall
 gravity theory proposed in 1972~\cite{Rastall:1973nw}. Rastall
 argued that the fundamental assumption $T^\nu{}_{\mu;\nu}=0$
 of GR is questionable in fact. All one can assert with fair
 confidence is~\cite{Rastall:1973nw}
 \be{eq8}
 T^\nu{}_{\mu;\nu}=X_\mu\,.
 \ee
 In~\cite{Rastall:1973nw}, Rastall proposed to consider the assumption
 \be{eq9}
 T^\nu{}_{\mu;\nu}=X_\mu=\lambda R_{,\mu}\,,
 \ee
 where $\lambda$ is a constant, and $R=R^\mu{}_\mu$. Since
 $G^\nu{}_{\mu;\nu}=0=\kappa\left(T^\nu{}_\mu-
 \lambda\,\delta^\nu{}_\mu R\right)_{;\nu}$ always, the
 assumption in Eq.~(\ref{eq9}) is consistent
 with the modified field equations~\cite{Rastall:1973nw}
 \be{eq10}
 G^\nu{}_\mu=\kappa\left(T^\nu{}_\mu-\lambda\,R\,\delta^\nu{}_\mu\right)
 \,,\quad\quad {\rm or\,\ equivalently}\quad\quad
 R_{\mu\nu}+\left(\kappa\lambda-\frac{1}{2}\right)g_{\mu\nu}R
 =\kappa T_{\mu\nu}\,,
 \ee
 where $\kappa$ is a non-zero constant. Contracting Eq.~(\ref{eq10})
 gives $\left(4\kappa\lambda-1\right)R=\kappa T$, where $T=T^\mu{}_\mu$,
 while the model parameters $\kappa\lambda=1/4$ should be excluded.
 Rastall~\cite{Rastall:1973nw} argued that
 \be{eq11}
 \frac{\kappa}{4\kappa\lambda
 -1}\left(3\kappa\lambda-\frac{1}{2}\right)=4\pi G_N\,.
 \ee
 When $\lambda=0$, we have $\kappa=8\pi G_N$, and GR is
 recovered. Obviously, Rastall gravity theory is a new kind
 of modified gravity theory violating the conservation of
 energy and momentum.

In the past decade, Rastall gravity was extensively studied in
 the literature (the original Ref.~\cite{Rastall:1973nw} has been cited
 more than 160 times to date), and we refer to~\cite{Capone:2009xm,
 Batista:2011nu,Fabris:2012hw,Silva:2012gn,Santos:2014ewa,Bronnikov:2016odv,
 Yuan:2016pkz,Heydarzade:2016zof,Heydarzade:2017wxu,Moradpour:2017shy,
 Spallucci:2017mto,Lobo:2017dib,Darabi:2017tay,Xu:2017bix,Kumar:2017qws,
 Visser:2017gpz,Moradpour:2017tbp,Salako:2018uzo,Lin:2018coh,Das:2018dzp,
 Abbas:2018ffk,Abbas:2018mxs,Bamba:2018zil,Sadeghi:2018vrf,Ziaie:2019jfl,
 Sakti:2019krw,Abbas:2019evn,Halder:2019akt,Khyllep:2019odd,Li:2019jkv} for
 example. In particular, very recently, it is claimed in~\cite{Li:2019jkv}
 that Rastall gravity is strongly favored by 118 galaxy-galaxy strong
 gravitational lensing systems, with $\kappa\lambda=0.163\pm 0.001$. This
 new observational evidence further justified the serious studies of
 Rastall gravity. In fact, the corresponding cosmology, black hole, compact
 star, wormhole, thermodynamics in Rastall gravity have been extensively
 considered in the literature. It attracted much attention in the recent
 years, and grows rapidly now.


\subsection{A novel modification to the Rastall gravity theory}\label{sec3b}

However, the original Rastall gravity theory is not suitable
 for our purpose. This can be clearly seen from
 Eq.~(\ref{eq10}). Since the last three diagonal
 components of the term $\lambda\,\delta^\nu{}_\mu R$ are
 equal, and the last three diagonal components of $T^\nu{}_\mu$
 are also equal for a perfect fluid, we must have
 $G^1{}_1=G^2{}_2=G^3{}_3$ again. Then, following the similar
 discussions in Sec.~\ref{sec2}, the LTB static universe in
 the original Rastall gravity theory also fails. So, we should
 introduce a novel modification to the Rastall gravity theory.

In Rastall gravity, Eq.~(\ref{eq8}) is firm, but the choice
 of $X_\mu$ might be changed. For simplicity, we
 propose to consider a fairly general form
 \be{eq12}
 T^\nu{}_{\mu;\nu}=X_\mu=Y^\nu{}_{\mu;\nu}\,,
 \ee
 where $Y^\nu{}_\mu\not=T^\nu{}_\mu$ (note that
 if $Y^\nu{}_\mu=\lambda\,R\,\delta^\nu{}_\mu$, the original
 Rastall gravity theory~\cite{Rastall:1973nw} can be recovered). In
 this new form, since
 $G^\nu{}_{\mu;\nu}=0=\kappa\left(T^\nu{}_\mu-Y^\nu{}_\mu\right)_{;\nu}$
 always, the assumption in Eq.~(\ref{eq12}) is consistent with
 the modified field equations
 \be{eq13}
 G^\nu{}_\mu+\kappa Y^\nu{}_\mu=\kappa T^\nu{}_\mu\,,
 \ee
 where $\kappa$ is a non-zero constant. Contracting Eq.~(\ref{eq13}), we
 find that the trace of $Y^\nu{}_\mu$ is given by
 \be{eq14}
 \kappa Y=R+\kappa T\,,
 \ee
 where $Y=Y^\mu{}_\mu$, $R=R^\mu{}_\mu$ and $T=T^\mu{}_\mu$. So far,
 the choice of $Y^\nu{}_\mu$ is still pending. Before we make a
 particular choice of $Y^\nu{}_\mu$, here are some general remarks:

\begin{enumerate}[(R1)]
\setlength{\itemindent}{0em}
  \item Regardless of the matter distribution in the universe,
  if $Y^\nu{}_\mu=0$, this Rastall-like gravity theory reduces to GR.
  \item If the matter distribution in the universe is isotropic and
  homogeneous, and if $Y^\nu{}_\mu$ is also isotropic and homogeneous
  (namely its last three diagonal components are equal and independent
  of spatial coordinates), the universe should be described by a FRW
  metric, and the Rastall-like gravity theory equivalently reduces to
  GR with an effective $T^\nu{}_{\mu,\,{\rm eff}}=T^\nu{}_\mu-Y^\nu{}_\mu$.
  \item If the matter distribution in the universe is isotropic and
  homogeneous, but $Y^\nu{}_\mu$ is anisotropic and/or inhomogeneous
  (namely its last three diagonal components are not equal and/or depend on
  spatial coordinates), because the anisotropic and/or inhomogeneous
  energy-momentum-exchange between matter and geometry will change the
  matter distribution in the universe, the universe becomes anisotropic
  and/or inhomogeneous (and hence is not described by a FRW metric).
  \item If the matter distribution in the universe is anisotropic and/or
  inhomogeneous, regardless of $Y^\nu{}_\mu$, the universe is
  not  described by a FRW metric.
\end{enumerate}

Next, let us step forward, and try to specify the choice of $Y^\nu{}_\mu$.
 We assume that the universe contains a perfect fluid whose
 energy-momentum tensor is given by Eqs.~(\ref{eq3}) or (\ref{eq4}).
 To obtain a successful LTB static universe, $Y^\nu{}_\mu$
 should satisfy three conditions:

\begin{enumerate}[(C1)]
\setlength{\itemindent}{0em}
  \item $Y^\nu{}_\mu$ is diagonal, because $G^\nu{}_\mu$ and $T^\nu{}_\mu$
  are both diagonal (n.b. Eqs.~(\ref{eq2}) and (\ref{eq4})).
  \item Its trace should be related to geometry and matter according
  to Eq.~(\ref{eq14}).
  \item Due to the discussions in Sec.~\ref{sec2},
  $Y^1{}_1\not=Y^2{}_2=Y^3{}_3$ is required to obtain a successful LTB
  static universe, while $G^1{}_1\not=G^2{}_2=G^3{}_3$ and
  $T^1{}_1=T^2{}_2=T^3{}_3$ (n.b. Eqs.~(\ref{eq2}) and (\ref{eq4})).
\end{enumerate}

\noindent Obviously, there is a large room for the choice of
 $Y^\nu{}_\mu$. The simplest one is given by
 \be{eq15}
 \kappa Y^\nu{}_\mu=\left(R+\kappa T\right){\cal J}^\nu{}_\mu\,,\quad\quad
 {\rm where}\quad\quad {\cal J}^\nu{}_\mu\equiv
 {\rm diag}\left(\,0,\,1,\,0,\,0\,\right)\,,
 \ee
 and we will briefly mention other reasonable choices in
 Sec.~\ref{sec6}. When $R+\kappa T=0$, this Rastall-like
 gravity theory reduces to GR (if $\kappa\not=8\pi G_N$, one
 can simply rescale $T^\nu{}_{\mu,\,{\rm new}}=
 \kappa T^\nu{}_\mu /(8\pi G_N)$, but $\kappa>0$ is required to
 ensure the energy density $\rho_{\rm new}\propto\kappa\rho\geq 0$). In
 some sense, our Rastall-like gravity is essentially different from GR and
 the original Rastall gravity (see the discussions in Sec.~\ref{sec6}).


\section{LTB static universe in Rastall-like gravity}\label{sec4}
In this section, we consider the LTB static universe in
 Rastall-like gravity. The field equations are given by
 Eq.~(\ref{eq13}), and $Y^\nu{}_\mu$ is given by
 Eq.~(\ref{eq15}). We assume that the universe contains
 a perfect fluid whose energy-momentum tensor is given by
 Eqs.~(\ref{eq3}) or (\ref{eq4}), and $p=w\rho$, where the
 equation-of-state parameter $w$ is a constant.

\vspace{-3.6mm} 


\subsection{LTB static solutions}\label{sec4a}

In the case of LTB static universe, $A$, $\rho$ and $p$ are
 all independent of the time $t$, and hence they are functions only
 depending on the spatial coordinate $r$, namely $A_0(r)$,
 $\rho_0(r)$ and $p_0(r)$, while we also denote $K=K_0(r)$.
 Due to the non-minimal coupling between geometry and matter,
 $T^\nu{}_{\mu;\nu}=Y^\nu{}_{\mu;\nu}\not=0$, one can find that
 $\rho_0^{\,\prime}\not=0$ and $p_0^{\,\prime}\not=0$, namely
 they are not homogeneous. The static solutions are determined by the
 field equations in Eq.~(\ref{eq13}), namely
 \bea
 -\frac{K_0}{A_0^2}-\frac{K_0^\prime}{A_0 A_0^\prime} &=& -\kappa\rho_0\,,
 \label{eq16}\\[0.6mm]
 -\frac{K_0}{A_0^2}+R_0+\kappa T_0 &=& \kappa p_0\,,\label{eq17}\\[-0.5mm]
 -\frac{K_0^\prime}{2 A_0 A_0^\prime} &=& \kappa p_0\,.\label{eq18}
 \eea
 Noting $R_0=2K_0/A_0^2+2K_0^\prime/(A_0 A_0^\prime)$ and
 $T_0=3p_0-\rho_0$, Eq.~(\ref{eq17}) is not independent of
 Eqs.~(\ref{eq16}) and (\ref{eq18}). Multiplying Eq.~(\ref{eq16}) by
 $w$, and then adding Eq.~(\ref{eq18}), we have
 \be{eq19}
 \frac{w K_0}{A_0^2}+
 \frac{K_0^\prime}{2A_0 A_0^\prime}\left(1+2w\right)=0\,.
 \ee
 Noting $K_0^\prime/A_0^\prime=dK_0/dA_0$, Eq.~(\ref{eq19}) can
 be regarded as an ordinary differential equation of $K_0$ with
 respect to $A_0$, and its solution is given by
 \be{eq20}
 K_0={\cal C}A_0^{-2w/(1+2w)}\,,
 \ee
 where $\cal C$ is an integral constant, and we require $w\not=-1/2$.
 Obviously, $K_0/A_0^2$ is not a constant if $w\not=-1/3$ and
 ${\cal C}\not=0$ (n.b. $A_0=A_0(r)$ varies with the spatial coordinate
 $r$), and hence the LTB static universe does not reduce to the FRW
 static universe. Multiplying Eq.~(\ref{eq18}) by $2$, and
 then subtracting Eq.~(\ref{eq16}), we have
 \be{eq21}
 \kappa\left(1+2w\right)\rho_0
 =\frac{K_0}{A_0^2}\,.
 \ee
 Substituting Eq.~(\ref{eq20}) into Eq.~(\ref{eq21}), it is easy to get
 \be{eq22}
 \rho_0=\frac{\cal C}{\kappa\left(1+2w\right)}\,A_0^{-2\,(1+3w)/(1+2w)}\,.
 \ee
 Eqs.~(\ref{eq20}) and (\ref{eq22}) are the explicit expressions of
 the LTB static solutions.


\subsection{Stability analysis}\label{sec4b}

To become a successful LTB static universe, it should be stable
 against perturbations. Fortunately, the comprehensive perturbation
 theory in LTB cosmology has been developed in~\cite{Clarkson:2009sc}.
 Because of the spherical symmetry of the LTB spacetime, perturbations
 can be decoupled into two independent modes, namely the polar and the
 axial modes~\cite{Clarkson:2009sc,Nishikawa:2014sga}. Since we are
 interested in the evolution of the density perturbations, we focus
 on the polar mode~\cite{Nishikawa:2014sga,February:2012fp}. Following
 e.g.~\cite{February:2012fp} and Sec.~III of \cite{Clarkson:2009sc},
 a first approximation is to neglect the mode-mixing, and
 focus only on the scalar perturbations~\cite{February:2012fp}. In the
 Regge-Wheeler (RW) gauge~\cite{Clarkson:2009sc,
 Nishikawa:2014sga,February:2012fp}, the perturbed metric is given by
 \be{eq23}
 g_{\mu\nu}=\bar{g}_{\mu\nu}+h_{\mu\nu}\,,
 \ee
 where $\bar{g}_{\mu\nu}$ is the background
 (static) metric, and~\cite{February:2012fp}
 \be{eq24}
 h_\mu{}^\nu={\rm diag}\left(-2\Phi,\,2\Psi,\,2\Psi,\,2\Psi\right)\,.
 \ee
 The perturbation of energy-momentum tensor is given by
 \be{eq25}
 \delta T_\mu{}^\nu=\left(\rho_0+p_0\right)\left(h_{\mu\sigma}U^\sigma U^\nu
 +\bar{g}_{\mu\sigma}U^\sigma \delta U^\nu
 +\bar{g}_{\mu\sigma}\delta U^\sigma U^\nu\right)
 +\left(\delta\rho+\delta p\right)\,\bar{g}_{\mu\sigma}U^\sigma U^\nu
 +\delta p\,\delta_\mu{}^\nu\,.
 \ee
 Note that the perturbations $\Phi$, $\Psi$, $\delta\rho$,
 $\delta p=w\delta\rho$, $\delta U^\mu$ are all functions of
 $t$, $r$, $\theta$, $\phi$. The perturbation of Ricci tensor
 induced by the perturbation of the metric is given by
 \be{eq26}
 \delta R_{\mu\sigma}=
 \frac{1}{2}\left(\nabla_\lambda\nabla_\mu h_\sigma{}^\lambda
 +\nabla_\lambda\nabla_\sigma h_\mu{}^\lambda
 -\nabla_\mu\nabla_\sigma h\right)
 -\frac{1}{2}\,g^{\alpha\beta}\nabla_\alpha\nabla_\beta h_{\mu\sigma}\,.
 \ee
 The perturbation of Ricci scalar reads
 \bea
 &\delta R&=g^{\mu\sigma}\delta R_{\mu\sigma}-h^{\mu\sigma}R_{\mu\sigma}
 =\nabla_\mu\nabla_\nu h^{\mu\nu}-\Box h -h^{\mu\nu}R_{\mu\nu}
 \nonumber\\[1.5mm]
 &&=2\ddot{\Psi}+2\ddot{\Phi}
 -\frac{4K_0}{A_0^2}\Psi-\frac{4K_0^\prime}{A_0 A_0^\prime}\Psi
 +2\,\Box\Psi-4\,\Box\Phi\,,\label{eq27}
 \eea
 where $\Box$ is the d'Alembertian. Substituting them into
 the perturbation of the field equation (\ref{eq13}), i.e.
 \be{eq28}
 \delta R_\mu{}^\nu-\frac{1}{2}\,\delta R\,\delta_\mu{}^\nu
 +\kappa\,\delta Y_\mu{}^\nu=\kappa\,\delta T_\mu{}^\nu\,,
 \ee
 its space-space ``$\,i\not=j\,$'' components tell us
 \be{eq29}
 \Phi=\Psi\,.
 \ee
 Substituting Eq.~(\ref{eq29}) into the diagonal components
 and the time-space ``$\,0\hspace{0.12em}i\,$'' components of
 Eq.~(\ref{eq28}), they become
 \bea
 4\ddot{\Psi}+\Box\Psi-\frac{1}{2}\,\delta R
 &=& -\kappa\,\delta\rho\,,\label{eq30}\\
 -\Box\Psi-\frac{2K_0^\prime}{A_0 A_0^\prime}\Psi+\frac{1}{2}\,\delta R
 +\kappa\left(3\delta p-\delta\rho\right) &=&
 \kappa\,\delta p\,,\label{eq31}\\[1.5mm]
 -\Box\Psi-2\left(\frac{K_0}{A_0^2}
 +\frac{K_0^\prime}{2A_0 A_0^\prime}\right)\Psi-\frac{1}{2}\,\delta R
 &=& \kappa\,\delta p\,,\label{eq32}\\[1.2mm]
 2\,\partial_i\,\partial_t\Psi &=&
 \left(\rho_0+p_0\right)\delta U_i\,.\label{eq33}
 \eea
 One can check that Eq.~(\ref{eq31}) is not independent of
 Eqs.~(\ref{eq30}) and (\ref{eq32}). Noting $g_{\mu\nu}U^\mu U^\nu=-1$, we
 have $\delta U_0=-h_{00}/2=-\Phi$. Substituting
 Eq.~(\ref{eq29}) into Eq.~(\ref{eq27}), it is easy to get
 \be{eq34}
 \delta R=4\ddot{\Psi}-\frac{4K_0}{A_0^2}\Psi
 -\frac{4K_0^\prime}{A_0 A_0^\prime}\Psi-2\,\Box\Psi\,.
 \ee
 Noting $\Box\Psi=-\ddot{\Psi}+\nabla^2\Psi$ in the case of LTB
 static universe ($\nabla^2$ is the Laplacian), and then
 substituting Eq.~(\ref{eq34}) into Eq.~(\ref{eq30}), we obtain
 \be{eq35}
 -\kappa\,\delta\rho=2\left(\frac{K_0}{A_0^2}+
 \frac{K_0^\prime}{A_0 A_0^\prime}\right)\Psi+
 2\,\nabla^2\Psi\,.
 \ee
 Once $\Psi$ is available, $\delta\rho$, $\delta U_i$ and $\delta U_0$ are
 ready by using Eqs.~(\ref{eq35}), (\ref{eq33}) and $\delta U_0=-\Psi$,
 respectively. If $\Psi$ is stable, they are also stable. Multiplying
 Eq.~(\ref{eq30}) by $w$, and then adding Eq.~(\ref{eq32}), we have
 \be{eq36}
 \ddot{\Psi}-\left[\left(\frac{K_0}{A_0^2}+
 \frac{K_0^\prime}{A_0 A_0^\prime}\right)w+
 \frac{K_0^\prime}{2A_0 A_0^\prime}\,\right]\Psi-w\nabla^2\Psi=0\,.
 \ee
 Using Eqs.~(\ref{eq16}) and (\ref{eq18}), it
 becomes
 \be{eq37}
 \ddot{\Psi}-w\nabla^2\Psi=0\,.
 \ee
 Considering a harmonic decomposition~\cite{Clarkson:2009sc} (see also
 e.g.~\cite{Li:2017ttl,Huang:2015kca,Harrison:1967zza}),
 \be{eq38}
 \Psi=\sum_{n=0}^\infty \psi_n(t)\,\Upsilon_n(r,\theta,\phi)=
 \sum_{n=0}^\infty \psi_n(t)\bigg[\,
 \xi_n(r)\sum_{m=-n}^n {\cal Y}_n^{\,m}(\theta,\phi)\,\bigg]\,,
 \ee
 Eq.~(\ref{eq37}) can be separated into two differential equations,
 namely
 \bea
 &\nabla^2 \Upsilon_n=-k^2 \Upsilon_n\,,\label{eq39}\\[1.5mm]
 &\ddot{\psi}_n+wk^2\psi_n=0\,,\label{eq40}
 \eea
 where the wave number $k$ is related to the
 degree $n$~\cite{Clarkson:2009sc,Harrison:1967zza,notek2}. One can obtain
 the spatial part of $\Psi$ by solving Eq.~(\ref{eq39})
 following e.g.~\cite{Clarkson:2009sc,Harrison:1967zza}, but it does not
 determine the stability of the perturbation $\Psi$. In fact, the stability
 of the perturbation $\Psi$ depends on the temporal part, $\psi_n(t)$, which
 can be stable on the condition $wk^2>0$ (n.b. Eq.~(\ref{eq40})). The LTB
 static universe is stable against the inhomogeneous ($k^2>0$) scalar
 perturbation if $w>0$. From Eq.~(\ref{eq21}), it is easy to see that $w>0$
 requires a closed ($K_0>0$) universe. Unfortunately, the LTB static
 universe is unstable against the homogeneous ($k^2=0$) scalar perturbation,
 since $\psi_n(t)\propto t$ diverges when $t\to\infty$. So, the LTB static
 universe fails in this case.


\section{LTB static universe in
 Rastall-like gravity with a~cosmological~constant}\label{sec5}

Let us come back to the discussions in Sec.~\ref{sec3b}. In fact, a
 cosmological constant $\Lambda$ can be allowed in the Rastall-like
 gravity theory. Since $\Lambda_{;\nu}=0$, it is easy to see
 that $\left(G^\nu{}_\mu+\Lambda\delta^\nu{}_\mu\right)_{;\nu}
 =0=\kappa\left(T^\nu{}_\mu-Y^\nu{}_\mu\right)_{;\nu}$ always holds. So,
 Eq.~(\ref{eq12}) is also consistent with the modified field equations
 \be{eq41}
 G^\nu{}_\mu+\Lambda\delta^\nu{}_\mu+\kappa Y^\nu{}_\mu
 =\kappa T^\nu{}_\mu\,.
 \ee
 Contracting Eq.~(\ref{eq41}), we find that the
 trace of $Y^\nu{}_\mu$ is given by
 \be{eq42}
 \kappa Y=R+\kappa T-4\Lambda\,.
 \ee
 The general remarks (R1)$\,\sim\,$(R4) in Sec.~\ref{sec3b} are
 still valid. To obtain a successful LTB static universe,
 the conditions (C1) and (C3) in Sec.~\ref{sec3b} are still
 valid, while the condition (C2) should be
 changed to Eq.~(\ref{eq42}). Again, there is a large room for
 the choice of $Y^\nu{}_\mu$. The simplest one is given by
 \be{eq43}
 \kappa Y^\nu{}_\mu
 =\left(R+\kappa T-4\Lambda\right){\cal J}^\nu{}_\mu\,,
 \ee
 where ${\cal J}^\nu{}_\mu$ is defined in Eq.~(\ref{eq15}),
 and we will briefly mention other reasonable choices in
 Sec.~\ref{sec6}. When $R+\kappa T-4\Lambda=0$,
 this Rastall-like gravity theory reduces
 to GR (if $\kappa\not=8\pi G_N$, one can simply rescale
 $T^\nu{}_{\mu,\,{\rm new}}=\kappa T^\nu{}_\mu /(8\pi G_N)$,
 but $\kappa>0$ is required to ensure the energy density
 $\rho_{\rm new}\propto\kappa\rho\geq 0$).

\vspace{-1.9mm} 


\subsection{LTB static solutions}\label{sec5a}

We assume that the universe contains a perfect fluid whose energy-momentum
 tensor is given by Eqs.~(\ref{eq3}) or (\ref{eq4}), and $p=w\rho$, where
 the equation-of-state parameter $w$ is a constant. In the case of LTB
 static universe, $A$, $\rho$ and $p$ are all independent of the time $t$,
 and hence they are functions only depending on the spatial coordinate $r$,
 namely $A_0(r)$, $\rho_0(r)$ and $p_0(r)$, while we also denote $K=K_0(r)$.
 Due to the non-minimal coupling between geometry and matter,
 $T^\nu{}_{\mu;\nu}=Y^\nu{}_{\mu;\nu}\not=0$, one can find that
 $\rho_0^{\,\prime}\not=0$ and $p_0^{\,\prime}\not=0$, namely they are not
 homogeneous. The static solutions are determined by the field equations
 in Eq.~(\ref{eq41}), namely
 \bea
 -\frac{K_0}{A_0^2}-\frac{K_0^\prime}{A_0 A_0^\prime}+\Lambda
 &=& -\kappa\rho_0\,,\label{eq44}\\[0.6mm]
 -\frac{K_0}{A_0^2}+R_0+\kappa T_0-3\Lambda &=&
 \kappa p_0\,,\label{eq45}\\[-0.5mm]
 -\frac{K_0^\prime}{2 A_0 A_0^\prime}+\Lambda
 &=& \kappa p_0\,.\label{eq46}
 \eea
 Noting $R_0=2K_0/A_0^2+2K_0^\prime/(A_0 A_0^\prime)$ and
 $T_0=3p_0-\rho_0$, Eq.~(\ref{eq45}) is not independent of
 Eqs.~(\ref{eq44}) and (\ref{eq46}). Multiplying Eq.~(\ref{eq44}) by
 $w$, and then adding Eq.~(\ref{eq46}), we have
 \be{eq47}
 \frac{w K_0}{A_0^2}+\frac{K_0^\prime}{2A_0 A_0^\prime}
 \left(1+2w\right)=\left(1+w\right)\Lambda\,.
 \ee
 Noting $K_0^\prime/A_0^\prime=dK_0/dA_0$, Eq.~(\ref{eq47}) can
 be regarded as an ordinary differential equation of $K_0$ with
 respect to $A_0$, and its solution is given by
 \be{eq48}
 K_0={\cal C}A_0^{-2w/(1+2w)}
 +\frac{\left(1+w\right)\Lambda}{1+3w}\,A_0^2\,,
 \ee
 where $\cal C$ is an integral constant, and we require $w\not=-1/2$
 and $w\not=-1/3$. Obviously, $K_0/A_0^2$ is not a constant if
 $w\not=-1/3$ and ${\cal C}\not=0$ (n.b. $A_0=A_0(r)$ varies with the
 spatial coordinate $r$), and hence the LTB static universe
 does not reduce to the FRW static universe. Multiplying
 Eq.~(\ref{eq46}) by $2$, and then subtracting Eq.~(\ref{eq44}), we have
 \be{eq49}
 \kappa\left(1+2w\right)\rho_0=\Lambda+\frac{K_0}{A_0^2}\,.
 \ee
 Substituting Eq.~(\ref{eq48}) into Eq.~(\ref{eq49}), it is easy to get
 \be{eq50}
 \rho_0=\frac{\cal C}{\kappa\left(1+2w\right)}\,A_0^{-2\,(1+3w)/(1+2w)}
 +\frac{2\Lambda}{\kappa\left(1+3w\right)}\,.
 \ee
 Eqs.~(\ref{eq48}) and (\ref{eq50}) are the explicit expressions of
 the LTB static solutions. Note that if $\Lambda=0$, all the
 results obtained here reduce to the ones in Sec.~\ref{sec4a}.
 But a non-zero $\Lambda$ makes difference.


\subsection{Stability analysis}\label{sec5b}

Again, to become a successful LTB static universe, it should be stable
 against perturbations. Similar to Sec.~\ref{sec4b}, we consider the
 perturbed metric given by Eqs.~(\ref{eq23}) and (\ref{eq24}).
 Accordingly, the perturbations
 $\delta T_\mu{}^\nu$, $\delta R_{\mu\sigma}$, $\delta R$ are
 the same given by Eqs.~(\ref{eq25})$\,\sim\,$(\ref{eq27}).
 Since $\delta\Lambda=0$, the perturbation of the field equation in
 Eq.~(\ref{eq41}) is still the same given in Eq.~(\ref{eq28}).
 Once again, its space-space ``$\,i\not=j\,$'' components tell
 us $\Phi=\Psi$ as in Eq.~(\ref{eq29}). Then, its diagonal components
 and the time-space ``$\,0\hspace{0.12em}i\,$'' components become the
 ones given in Eqs.~(\ref{eq30})$\,\sim\,$(\ref{eq33}).
 Of course, Eqs.~(\ref{eq34})$\,\sim\,$(\ref{eq36}) still hold.
 However, the static solutions in Eqs.~(\ref{eq44}) and (\ref{eq46})
 make difference. Substituting
 Eqs.~(\ref{eq44}) and (\ref{eq46}) into Eq.~(\ref{eq36}), it becomes
 \be{eq51}
 \ddot{\Psi}-\left(1+w\right)\Lambda\Psi-w\nabla^2\Psi=0\,,
 \ee
 which is different from Eq.~(\ref{eq37}) if $\Lambda\not=0$
 and $w\not=-1$. Again, considering the harmonic decomposition given
 in Eq.~(\ref{eq38}), we can separate Eq.~(\ref{eq51}) into
 two differential equations, namely
 \bea
 &\nabla^2 \Upsilon_n=-k^2 \Upsilon_n\,,\label{eq52}\\[1.5mm]
 &\ddot{\psi}_n+\left[\,wk^2-\left(1+w\right)\Lambda\,\right]\psi_n
 =0\,,\label{eq53}
 \eea
 where the wave number $k$ is related to the
 degree $n$~\cite{Clarkson:2009sc,Harrison:1967zza,notek2}.
 One can obtain the spatial part of $\Psi$ by solving Eq.~(\ref{eq52})
 following e.g.~\cite{Clarkson:2009sc,Harrison:1967zza}, but it does
 not determine the stability of the perturbation $\Psi$. In fact, the
 stability of the perturbation $\Psi$ depends on the temporal part,
 $\psi_n(t)$. From Eq.~(\ref{eq53}), it is easy to see that the
 stability condition for the LTB static universe reads
 \be{eq54}
 wk^2-\left(1+w\right)\Lambda>0\,.
 \ee
 The LTB static universe is stable against the homogeneous ($k^2=0$)
 scalar perturbation if
 \be{eq55}
 \left(1+w\right)\Lambda<0\,.
 \ee
 On the other hand, the LTB static universe is also stable
 against the inhomogeneous ($k^2>0$) scalar perturbation if
 Eq.~(\ref{eq54}) is satisfied for all possible modes with
 $k^2>0$. To ensure that Eq.~(\ref{eq54}) is valid even when
 $k^2\to\infty$, it is necessary to require
 \be{eq56}
 w\geq 0\,.
 \ee
 Combining Eqs.~(\ref{eq55}) and (\ref{eq56}), the LTB static
 universe can be stable against both the homogeneous ($k^2=0$)
 and the inhomogeneous ($k^2>0$) scalar perturbations if
 \be{eq57}
 w\geq 0
 \quad\quad {\rm and} \quad\quad \Lambda<0\,.
 \ee
 From Eq.~(\ref{eq49}), we find that $w\geq 0$ and $\Lambda<0$
 require a closed ($K_0>0$) universe. So far, we successfully
 get a stable LTB closed static universe on the conditions given by
 Eq.~(\ref{eq57}).


\section{Concluding remarks}\label{sec6}

In this work, we try to obtain a stable LTB static universe, which is
 spherically symmetric and radially inhomogeneous. However, this is not an
 easy task, and fails in GR and various modified gravity theories, because
 the corresponding LTB static universes must reduce to the FRW static
 universes. We find a way out in a new type of modified gravity theory,
 in which the conservation of energy and momentum is broken. In this work,
 we have proposed a novel modification to the original Rastall gravity. In
 some sense, our Rastall-like gravity is essentially different from GR and
 the original Rastall gravity (see below). In this Rastall-like gravity, LTB
 static solutions have been found. The stability of LTB static universe
 against both the homogeneous and the inhomogeneous scalar perturbations is
 also discussed in details. We show that a LTB static universe can be stable
 in this Rastall-like gravity.

As is well known, the stability conditions for many FRW
 (Einstein) static universes in various modified gravity theories are
 fairly complicated, and usually require exotic matter with $w<0$, in
 particular dark energy ($w<-1/3$) or even phantom ($w<-1$). On the
 contrary, the stability conditions for the LTB static universe given
 in Eq.~(\ref{eq57}) is very simple. Obviously, the condition
 $w\geq 0$ can be easily satisfied by using ordinary matter, such as
 radiation ($w=1/3$) or dust matter ($w=0$). On the other hand,
 a negative cosmological constant ($\Lambda<0$) is also welcome
 in e.g. string theory. Although the current accelerated expansion of
 the universe requires dark energy ($w<-1/3$) or a positive
 cosmological constant ($\Lambda>0$), this is not the case of
 LTB static universe (as the initial state for the past-eternal early
 universe in the emergent universe scenario).

It is worth noting that the (effective) gravitational forces
 provided by the ordinary matter ($w\geq 0$) and a negative
 cosmological constant ($\Lambda<0$) are attractive. This means
 that the effective force contributed by the non-minimal
 coupling between matter and geometry $Y^\nu{}_{\mu;\nu}$ is
 repulsive, which can also be seen from the first equation of
 $T^\nu{}_{\mu;\nu}=X_\mu=Y^\nu{}_{\mu;\nu}$, namely
 \be{eq58}
 \dot{\rho}+\left(2\,\frac{\dot{A}}{A}+\frac{\dot{A}^\prime}{A^\prime}
 \right)\left(\rho+p+p_{\rm eff}\right)=0\,,
 \ee
 with a negative effective pressure $p_{\rm eff}<0$ coming
 from $X_\mu=Y^\nu{}_{\mu;\nu}$ (note that Eq.~(\ref{eq58}) in
 the LTB cosmology corresponds to the familiar
 $\dot{\rho}+3H\left(\rho+p+p_{\rm eff}\right)=0$ in the FRW
 cosmology). So, when the (effective) attractive forces
 are stably balanced by the effective repulsive force, the LTB static
 universe can be accomplished.

In this work, we assume that the universe contains a perfect fluid.
 Actually, one can also extend our discussions to a non-perfect
 fluid, for example, van der Waals fluid, viscous fluid, and
 Newtonian fluid. Of course, the role of matter can also be played by
 a scalar field or a vector field. In Rastall-like gravity, we expect
 that a stable LTB static universe can also be accomplished in these cases.

For simplicity, in this work we only consider the simplest choices of
 $Y^\nu{}_\mu$, as given in Eqs.~(\ref{eq15}) and (\ref{eq43}).
 Actually, there is a large room for other reasonable choices.
 In the case without a cosmological constant, one might instead
 consider, for example,
 \be{eq59}
 \kappa Y^\nu{}_\mu={\rm diag}\left(\,R,\,\kappa T,\,0,\,0\,\right)\,,
 \quad\quad {\rm or} \quad\quad
 \kappa Y^\nu{}_\mu={\rm diag}\left(\,\kappa T,\,R,\,0,\,0\,\right)\,,
 \ee
 or even a more general choice
 \be{eq60}
 \kappa Y^\nu{}_\mu={\rm diag}\left(\,\alpha R+\beta\kappa T,\,
 \left(1-\alpha\right)R+\left(1-\beta\right)\kappa T,\,0,\,0\,\right)\,,
 \ee
 where $\alpha$ and $\beta$ are both constants. Further, if we are willing
 to involve the other two spatial components, it is also possible to choose
 \be{eq61}
 \kappa Y^\nu{}_\mu={\rm diag}\left(\,\left(1-\alpha_1-2\alpha_2\right)R
 +\left(1-\beta_1-2\beta_2\right)\kappa T,\,\alpha_1 R+\beta_1\kappa T,\,
 \alpha_2 R+\beta_2\kappa T,\,\alpha_2 R+\beta_2\kappa T\,\right)\,,
 \ee
 where $\alpha_i$ and $\beta_i$ are arbitrary constants. Obviously,
 Eqs.~(\ref{eq59}) and (\ref{eq60}) are just special cases of
 Eq.~(\ref{eq61}). It is worth noting that the general
 $\kappa Y^\nu{}_\mu$ in Eq.~(\ref{eq61}) can be recast as
 \bea
 &\kappa Y^\nu{}_\mu &= \frac{1}{2}\,\big[\left(1-\alpha_1-\alpha_2\right)R
 +\left(1-\beta_1-\beta_2\right)\kappa T\,\big]\,\delta^\nu{}_\mu\nonumber
 \\[1.2mm]
 & & -\,\frac{1}{2}\,\big[\left(1-\alpha_1-3\alpha_2\right)R+
 \left(1-\beta_1-3\beta_2\right)\kappa T\,\big]\,\eta^\nu{}_\mu
 \nonumber\\[1.5mm]
 & & +\,\big[\left(\alpha_1-\alpha_2
 \right)R+\left(\beta_1-\beta_2\right)\kappa T\,\big]\,{\cal J}^\nu{}_\mu
 \,,\label{eq62}
 \eea
 where $\eta^\nu{}_\mu={\rm diag}\left(-1,\,1,\,1,\,1\right)$,
 and ${\cal J}^\nu{}_\mu$ is defined in Eq.~(\ref{eq15}). Of course,
 in the case with a cosmological constant $\Lambda$, the choices are
 quite similar, while one should appropriately insert $\Lambda$ into
 Eqs.~(\ref{eq59})$\,\sim\,$(\ref{eq62}). Note that all the choices
 mentioned above satisfy the conditions (C1)$\,\sim\,$(C3)
 to obtain a successful LTB static universe, as in Sec.~\ref{sec3b}
 or Sec.~\ref{sec5}. However, when we consider other topics rather than
 static universe, the conditions (C1) and (C3) could be abandoned, and we
 can then adopt other suitable choices of $Y^\nu{}_\mu$, while the trace
 condition (C2) is still required.

It is of interest to discuss the key difference between our Rastall-like
 gravity proposed in this work and the original Rastall gravity proposed
 in~\cite{Rastall:1973nw}. Actually, it is argued
 in e.g.~\cite{Visser:2017gpz} that one can alternatively regard the
 original Rastall gravity proposed in~\cite{Rastall:1973nw} as GR with
 a hypothetical matter sector. Contracting Eq.~(\ref{eq10})
 gives $\left(4\kappa\lambda-1\right)R=\kappa T$. Then, the field equations
 of the original Rastall gravity, namely Eq.~(\ref{eq10}), could be recast
 as (see e.g.~\cite{Batista:2011nu})
 \be{eq63}
 G^\nu{}_\mu=\kappa\tilde{T}^\nu{}_\mu\,,\quad\quad {\rm or\,\ equivalently}
 \quad\quad R_{\mu\nu}-\frac{1}{2}\,g_{\mu\nu}R=\kappa\tilde{T}_{\mu\nu}\,,
 \ee
 where the effective energy-momentum tensor is given by
 \be{eq64}
 \tilde{T}^\nu{}_\mu=T^\nu{}_\mu-
 \frac{\kappa\lambda}{4\kappa\lambda-1}\,\delta^\nu{}_\mu T\,,
 \quad\quad {\rm or\,\ equivalently} \quad\quad
 \tilde{T}_{\mu\nu}=T_{\mu\nu}-
 \frac{\kappa\lambda}{4\kappa\lambda-1}\,g_{\mu\nu}T\,,
 \ee
 in which we have used the relation $\left(4\kappa\lambda-1\right)R=\kappa T$.
 Note that $\tilde{T}^\nu{}_\mu$ is completely determined by the usual
 matter $T^\nu{}_\mu$ (n.b.~Eq.~(\ref{eq64}) and $T=T^\mu{}_\mu$). This is
 the key point of e.g.~\cite{Visser:2017gpz} arguing that the original
 Rastall gravity proposed in~\cite{Rastall:1973nw} is equivalent to GR with
 a hypothetical matter sector. However, in e.g.~\cite{Moradpour:2017tbp},
 it is argued that the same logic can also be applied to any modified
 gravity theory. In fact, one can always recast the field equations of
 modified gravity theory as the form of Eq.~(\ref{eq7}), where
 $M^\nu{}_\mu$ is the modification term with respect to GR. Similarly,
 the field equations of any modified gravity theory, namely
 Eq.~(\ref{eq7}), could also be recast as
 \be{eq65}
 G^\nu{}_\mu=8\pi G_N\tilde{T}^\nu{}_\mu\,,\quad\quad {\rm where}\quad\quad
 \tilde{T}^\nu{}_\mu=T^\nu{}_\mu-M^\nu{}_\mu/\left(8\pi G_N\right)\,,
 \ee
 which has the same form of GR. However, as is well known, most of the
 modified gravity theories in the literature (e.g.~$f(R)$ theories)
 are essentially different from GR, and have richer phenomena than GR.
 They are not equivalent to GR in fact. Therefore, the arguments of
 e.g.~\cite{Visser:2017gpz} are disagreed by e.g.~\cite{Moradpour:2017tbp}.
 Here, let us push these discussions further. Actually, in most of the
 modified gravity theories in the literature, the modification term
 $M^\nu{}_\mu$ is a function of geometric quantities. For example, in
 the well-known $f(R)$ theories (we adopt the
 form of ${\cal S}_{\rm grav}\propto\int d^4 x\sqrt{-g}\left(R+f(R)\right)$
 in the metric formalism), the modification term with respect to GR
 is given by (see e.g.~\cite{Seahra:2009ft})
 \be{eq66}
 M_{\mu\nu}=f_{,R\,}R_{\mu\nu}-\frac{1}{2}fg_{\mu\nu}+
 \left(\,g_{\mu\nu}\Box-\nabla_\mu\nabla_\nu\right)f_{,R}\,,
 \ee
 where $f_{,R}=df/dR$. Clearly, it is essentially a geometric quantity.
 So, there exists a key difference between the original Rastall gravity
 proposed in~\cite{Rastall:1973nw} and most of the modified gravity
 theories in the literature. If the usual mater sector $T^\nu{}_\mu$ is
 given, in the original Rastall gravity~\cite{Rastall:1973nw},
 the corresponding $\tilde{T}^\nu{}_\mu$ in Eq.~(\ref{eq64}) is
 completely determined, and it is a matter quantity essentially. However,
 this is not the case in most of the modified gravity theories in the
 literature (e.g.~$f(R)$ theories), because $M_{\mu\nu}$ is a geometric
 quantity, or $M_{\mu\nu}$ depends on both the geometric and the matter
 sectors, and hence the corresponding $\tilde{T}^\nu{}_\mu$ in
 Eq.~(\ref{eq65}) is not a pure matter quantity. Even if the usual mater
 sector $T^\nu{}_\mu$ is given, this $\tilde{T}^\nu{}_\mu$
 still cannot be explicitly determined. In this sense, most of
 the modified gravity theories in the literature (e.g.~$f(R)$ theories)
 are essentially different from GR and the original Rastall gravity
 theory~\cite{Rastall:1973nw}. We hope this insight could reconcile
 the disputation between e.g.~\cite{Visser:2017gpz} and
 \cite{Moradpour:2017tbp}. Now, let us turn to our Rastall-like gravity
 proposed in this work. Clearly, our $M^\nu{}_\mu=\kappa Y^\nu{}_\mu$
 (n.b. Eq.~(\ref{eq13})) depends on both the geometric and the matter
 sectors. Actually, in all choices given in Eqs.~(\ref{eq15}),
 (\ref{eq43}), and (\ref{eq59})$\,\sim\,$(\ref{eq62}), the corresponding
 $M^\nu{}_\mu=\kappa Y^\nu{}_\mu$ depend on both $R$ and $T$. Unlike
 the original Rastall gravity proposed in~\cite{Rastall:1973nw}, the
 relation $\left(4\kappa\lambda-1\right)R=\kappa T$ is not valid in
 our Rastall-like gravity. Due to the lack of the explicit relation
 between $R$ and $T$, our $M^\nu{}_\mu=\kappa Y^\nu{}_\mu$ certainly
 cannot be a pure matter quantity, and at least part of
 $M^\nu{}_\mu=\kappa Y^\nu{}_\mu$ comes from geometry. Even if the usual
 mater sector $T^\nu{}_\mu$ is given, the corresponding
 $\tilde{T}^\nu{}_\mu$ in Eq.~(\ref{eq65}) still cannot be explicitly
 determined. In this sense, our Rastall-like gravity proposed in this work
 is essentially different from GR and the original Rastall gravity proposed
 in~\cite{Rastall:1973nw}, and actually it is somewhat similar to most of
 the modified gravity theories in the literature (e.g. $f(R)$ theories).
 Without the need of exotic matter, the modification term coming from
 geometry makes difference, as in the cases of other modified
 gravity theories in the literature (e.g. $f(R)$ theories).

In a successful emergent universe scenario, the universe must exit the
 static state and then enter an expansion phase (we thank the referee for
 pointing out this issue). For example, in Eddington-inspired Born-Infeld
 (EiBI) theory~\cite{Banados:2010ix}, the universe can be past-eternal
 (see~\cite{Li:2017ttl} for stability analysis), and then expands in the
 late time (see the bottom panel ($\kappa>0$) of Fig.~2
 in~\cite{Banados:2010ix}). Noting that the ultimate theory of gravity
 is not available so far, the existing theories (including our Rastall-like
 gravity theory) might be just approximations of a fundamental theory. In
 the past ($t\to -\infty$), the difference between Rastall-like gravity
 theory and the fundamental theory can be negligible, and hence the static
 universe is past-eternal. However, in the late time, the difference
 between Rastall-like gravity theory and the fundamental theory becomes
 significant. In this case, the universe accordingly becomes unstable in
 the presence of perturbations. Then, the universe exits the static state
 and enters an expansion phase.


\section*{ACKNOWLEDGMENTS}
We thank the anonymous referee for useful comments and suggestions,
 which helped us to improve this work. We are grateful to Zhao-Yu~Yin,
 Da-Chun~Qiang, Hua-Kai~Deng and Shu-Ling~Li for kind help and discussions.
 This work was supported in part by NSFC under Grants No.~11975046 and
 No.~11575022.

\renewcommand{\baselinestretch}{1.024}


\end{document}